\documentclass[aps,pre,amsfonts,showpacs,unsortedaddress]{revtex4}

\usepackage[english]{babel}
\usepackage[latin1]{inputenc}
\usepackage[dvips]{graphicx}
\usepackage{amsmath}
\usepackage{amssymb}
\usepackage{epsfig}
\usepackage{array}

\newcommand{\be}{\begin{equation}}
\newcommand{\ee}{\end{equation}}
\newcommand{\ba}{\begin{eqnarray}}
\newcommand{\ea}{\end{eqnarray}}
\newcommand{\nn}{\nonumber\\}
\newcommand{\Ref}[1]{(\ref{#1})}
\newcommand{\av}[1]{\langle #1\rangle}
\newcommand{\half}{\textstyle{\frac{1}{2}}}
\newcommand{\br}{{\bf r}}
\newcommand{\bv}{{\bf v}}
\newcommand{\bk}{{\bf k}}
\newcommand{\bX}{{\bf X}}
\newcommand{\bnabla}{\mbox{\boldmath$\nabla$}}
\newcommand{\fourth} {\textstyle{\frac{1}{4}}}

\begin{document}

\title{Model system for classical fluids out of equilibrium}
\author{M. Ripoll$^{a,b}$\thanks{Author to whom correspondence
should be addressed, e-mail:mripoll@fz-juelich.de} and M. H.
Ernst$^{c,d}$}
\affiliation{$(a)$ Institut f\"ur Festk\"orperforschung,
Forschungszentrum J\"ulich -  52425 J\"ulich, Germany\\
$(b)$ Dpto F\'{\i}sica Fundamental,
UNED, C/Senda del Rey 9, 28040 Madrid, Spain \\
$(c)$ CNLS, Los Alamos National
Laboratory, Los Alamos, NM 87545, USA \\
$(d)$ Institute for Theoretical Physics, Utrecht University,
Princetonplein 5, P.O. Box 80.195,  3508 TD  Utrecht, The
Netherlands}

\date{\today}

\begin{abstract}
A model system for classical fluids out of equilibrium, referred
to as DPD solid (Dissipative Particles  Dynamics), is studied by
analytical and simulation methods. The time evolution of a DPD
particle is described by a fluctuating heat equation. This DPD
solid with transport based on collisional transfer (high density
mechanism) is complementary to the Lorentz gas with only kinetic
transport (low density mechanism). Combination of both models
covers the qualitative behavior of transport properties of
classical fluids over the full density range. The heat diffusivity
is calculated using a mean field theory, leading to a linear
density dependence of this transport coefficient, which is exact
at high densities. Subleading density corrections are obtained as
well. At lower densities the model has a conductivity threshold
below which heat conduction is absent. The observed threshold is
explained in terms of percolation diffusion on a random proximity
network. The geometrical structure of this network is the same as
in continuum percolation of completely overlapping spheres, but
the dynamics on this network differs from continuum percolation
diffusion. Furthermore, the kinetic theory for DPD is extended to
the generalized hydrodynamic regime, where the wave number
dependent decay rates of the Fourier modes of the energy and
temperature fields are calculated. \\
\end{abstract}

\maketitle

\renewcommand{\theequation}{I.\arabic{equation}}
\setcounter{section}{0} \setcounter{equation}{0}
\section{Introduction}

The Lorentz gas, describing the self diffusion of a moving
particle in a random array of scatterers, has played an important
role in understanding the transport properties of classical
fluids, and in developing and quantifying the role of correlated
sequences of binary collisions (ring collisions), as well as in
extending the kinetic theory to moderately dense fluids (see
\cite{vLW67,EW79,Alder-alley,Machta,machta-moore85} and references
there in).

 The Lorentz gas contains only the mechanism of {\it kinetic}
transport, which is the most important transport mechanism at low
densities. In dense fluids there is a second mechanism for
transport, called { \it collisional transfer} \cite{chapman},
through which energy and momentum is instantaneously transferred
through the interactions between particles  within each others
force range. This is the dominant mechanism at high densities.

In this paper we discuss a model, {\it complementary} to the Lorentz
gas, which contains only the mechanism of collisional transfer, and
for which transport properties can be evaluated exactly in the high
density limit. It is also important to develop systematic theories for
subleading large density corrections. The combination of both
complementary models, the Lorentz gas and the DPD solid, might be able
to describe the qualitative density dependence of transport properties
of classical fluids over the full range from low to high densities.

Before introducing the random DPD solid we briefly discuss the
relevance of DPD models for the study of classical fluids. This
stems from the great interest in computer simulations of complex
fluids and quenched random media, which are challenging problems
as several different space and time scales may be involved. Fully
atomistic simulations of such systems fail in reaching the larger
scales, and different mesoscopic models/techniques, such as
Dissipative Particle Dynamics (DPD), Smooth Particle
Hydrodynamics, cellular automata and lattice gases, lattice
Boltzmann methods, etc, offer possibilities to explore these
larger scales. For that reason the DPD technique was originally
introduced \cite{hoo92} as a mesoscopic particle method for
simulating complex fluids and colloidal suspensions. It is
therefore important to provide a theoretical analysis of such
systems, as will be done in this paper.

The idea of the method is that each DPD (point) particle
represents a mesoscopic portion of fluid. The interactions among
these point particles have no hard core, and are softly repulsive.
The lack of hard core interactions allows time-step driven
algorithms \cite{PHF98}. In the original formulation the DPD
particle is described in terms of its position and velocity with
three different types of interactions: conservative, dissipative
and stochastic. The forces between particles are pair-wise, such
that mass and momentum are conserved, but energy is not. This
formulation is restricted to isothermal problems, but describes a
proper hydrodynamic behavior for viscous fluids
\cite{EW95,mbe1,PHF98,MW99} in a large number of problems. The
method has also been successful  in describing  properties of
colloidal suspensions \cite{boe97,dz00a}, polymer solutions
\cite{kong97,spen00}, phase separation \cite{wij01,cov96} or
membranes \cite{smit03,groot01}.  In standard DPD the forces are
Gaussian white noise, which have been recently extended to colored
noise \cite{Reich}.

A generalization of the model to include energy conservation has
been proposed as well \cite{esp97e,bonet1} in order to describe
heat flows and other thermal effects in fluids out of equilibrium.
In the picture where DPD particles are understood as droplets or
mesoscopic clusters of microscopic particles, one can consider the
kinetic energy lost in dissipative interactions as being
transformed into energy of internal degrees of freedom of a
particle. The number of internal states of a DPD particle with
energy $\epsilon$, $\exp [s(\epsilon)/k_B]$, is modelled by an
entropy function $s(\epsilon)$, which implies a temperature
$T(\epsilon)$ defined through $\partial s(\epsilon) / \partial
\epsilon = 1 / T(\epsilon)$. The evolution of the internal energy
has two contributions. The first one describes how the friction
forces contribute to the change of kinetic energy; this is {\em
viscous heating}. In addition, the phenomenon of {\em heat
conduction} has to be considered, where  energy or temperature
differences between particles (subsystems) produce a flux of
internal energy. In Ref. \cite{bonet2} Bonet and Mackie have
estimated the transport coefficients in the limit of high friction
for this thermal DPD fluid, using a method somewhat similar to
that used in Ref. \cite{hoo92}. In Ref. \cite{Ripoll-thesis} we
have investigated the DPD fluid obeying the conservation laws of
mass, momentum and energy, and we have calculated the full set of
transport coefficients in the Navier-Stokes and energy balance
equation, including kinetic and collisional transfer
contributions, as well as their wave length dependent
generalizations. However, as pointed out in Ref.\cite{PHF98} for
standard DPD fluids, the theoretical values for the transport
coefficients appear to agree only well with computer simulations
at higher densities.

To analyze the difficulties and to develop a more accurate
description it is of interest to consider a simplified model with
heat conduction, the {\it random DPD solid}, which still has the
basic features of DPD. Such a model can be obtain by considering
the energy conserving DPD models \cite{esp97e,bonet1}, where the
dynamical degrees of freedom of a particle, $x_i(t) = \{\br_i(t),
\bv_i(t), \epsilon_i(t) \}$ $ (i=1,2, \cdots,N)$ are position,
velocity and internal energy. By {\it freezing/quenching} the
velocities, the particles can be characterized  by  {\it static}
(random) positions $ \br_i$, and by {\it dynamic} energy variables
$\epsilon_i(t)$, with their total energy, $E=\sum_i
\epsilon_i(t)$, conserved.

Consequently, the {\it macroscopic} evolution equation for this
system is Fourier's law of heat diffusion, and the system is able
to carry a macroscopic heat flux, provided a {\it macroscopic
fraction} of the particles is inside each other's interaction
ranges. For simplicity we set the conservative forces equal to
zero (point particles), and take the interaction range of the
dissipative and stochastic forces equal to $r_c$. This model has
been proposed in \cite{ripoll98}, basically as a {\it discrete
fluctuating heat equation}.

Transport of energy in fluids consists in general of {\it kinetic}
transport, carried by moving particles, and instantaneous
transport through the interactions, the so-called {\it collisional
transfer}. In the present model there is no kinetic transport, and
the only type of transport is through collisional transfer. In
dense fluids collisional transfer is also the dominant mechanism
of transport.

The basic observation is   that the collisional transfer mechanism
can be viewed as hopping of energy across existing {\it bonds}, as
illustrated directly by the dissipative part of the $N-$particle
Langevin equations, i.e.,
\be \label{langevin-no-noise}
d\epsilon_i /dt = \lambda_0 {\sum_j}^\prime w(r_{ij}) (\epsilon_i
- \epsilon_j).
\ee
It is essentially a discrete diffusion equation on a random
network with bonds, to be defined below. In the previous equation
$\lambda_0$ is a relaxation coefficient, and $w(r_{ij})$ is a
positive interaction function, say, equal to 1 for $r_{ij} \leq
r_c$  and 0 elsewhere, where $r_c$ is the interaction range. Here
we are dealing with a dynamic diffusion problem on an underlying
static percolating structure. In order to have any transport of
energy  the density should be sufficiently large, such that there
exists a percolating path of {\it connected} particles, spanning
two opposite boundaries of the system. Groups of isolated or
connected particles, which are not part of the percolating path,
form non-conducting islands. So the underlying static structure is
a {\it bond-percolation} cluster, whose density should not be
below some threshold density \cite{stauffer-aharony}.

The question is then, what is a {\it bond} in this quenched random
solid of point particles occupying {\it random} positions
$\{\br_i| i=1,2,\cdots, N\}$ and having an interaction range
$r_c$. To visualize the connected network, we connect every pair
$\{i,j\}$ of fixed point particles with a bond if $|\br_{ij}| \leq
r_c$. Energy can only hop between connected particles. With this
definition of a bond, we have a well defined  percolation
diffusion model on a random {\it proximity} network with a {\it
constant} hopping rate per bond.

The geometrical structure of this connected network is the same as
in continuum percolation of completely overlapping spheres ( see
Ref. \cite{ziff-3d,ziff} and references therein). Suppose we put
black circles of radius $R= \half r_c$ on every point particle.
Then two particles $\{i,j\}$ are considered to be connected or
"overlapping" if $r_{ij} \leq 2R=r_c$, and we obtain the above
continuum percolating structure.

However the dynamics of the present diffusion problem is quite
different from continuum percolation diffusion models, such as the
overlapping Lorentz gas \cite{machta-moore85} and the Swiss Cheese
model, where diffusion occurs in the void spaces {\it outside} the
overlapping spheres \cite{feng-halperin87,stenull01}, or the
Inverted Swiss Cheese model and others \cite{feng-halperin87},
where diffusion occurs in the complementary space, i.e. {\it
inside} the overlapping spheres. Such models can be mapped either
on the discrete random proximity network (Inverted Swiss Cheese
model), or on its {\it dual network} (overlapping Lorentz gas,
Swiss Cheese model). The big difference is that the maps of the
continuum diffusion models have a wide distribution of hopping
rates, usually singular at small rates, because of the appearance
of bottleneck passages \cite{kerst83}, whereas the random DPD
solid has {\it constant} or nearly constant  hopping rates
(depending on the shape of the range function $w(r)$).

We also note that the overlapping Lorentz gas and the Swiss Cheese
model are percolating below a threshold density, whereas the  {\it
dual } models:  the Inverted Swiss Cheese model and the random DPD
solid are percolating above a threshold density, illustrating the
relevance of these models for low or high density fluids.

In the overlapping Lorentz gas, say in two-dimensions, the
disordered network is formed by the vertices and edges ("bonds")
of the polygons that partition space into a Voronoi tessellation
\cite{machta-moore85,feng-halperin87}. Here the {\it blocked
bonds} are the edges that are perpendicular to and bisect the
lines of centers {\it with} a length $r_{ij} \leq r_c$. Hence, the
blocked bonds in the Voronoi tessellation are the {\it duals} of
the bonds in the random proximity network. Dynamical properties
(exponents, amplitudes) near the threshold density will in general
be different on {\it discrete} disordered networks with constant
hopping rates, such as the random DPD solid, and on {\it
continuum} percolation models, corresponding to networks with a
wide (singular) distribution of hopping rates.

From the point of view of dense fluids the percolation phenomena
are in a way just an interesting pathology of the random DPD
solid, caused by freezing out the translational degrees of freedom
of the corresponding DPD fluid. So the main interest of this paper
is focused on a {\it quantitative} description of transport
coefficients {\it away } from the percolation density. This
kinetic theory problem has not received much attention  in the
literature of the last three decades, which has been focusing on
the behavior near the threshold, and not on the density dependence
away from the threshold density.

So far we have described the dissipative part of the fluctuating
heat equation. An $N-$particle state described by the dissipative
equation \Ref{langevin-no-noise} would decay from an arbitrary
initial state to a state of {\it zero temperature} with all
energies $\epsilon_i =E/N$. To make the DPD solid reach
thermodynamic equilibrium one adds a fluctuating force to the
evolution equation, satisfying the fluctuation-dissipation
theorem. How this is done is explained in Section 2 and Appendix
A.

The plan of the paper is as follows. It starts with a more
detailed presentation of the heat conducting random DPD solid in
Section II, while some more general properties, such as the ${\cal
H}$ theorem and the equilibrium properties are discussed in
Appendix A. In Section III A the heat diffusivity is calculated
using a mean field theory which is expected to be exact at large
densities, where local density fluctuations can be neglected
\cite{mbe1,rip01}. In Section III B the large deviations at low
densities between the results of mean field theory and simulation
are explained in terms of bond percolation on a random proximity
network. In Section IV  we derive the wave number dependent decay
rates of the spatial Fourier modes of the fluctuations in energy
and temperature fields. The last section is devoted to some
conclusions.

\renewcommand{\theequation}{II.\arabic{equation}}
\setcounter{section}{1} \setcounter{equation}{0}
\section{DPD - Heat conduction model}

The heat conduction in dissipative particle dynamics is modelled
as a thermally conducting {\it quenched random} solid. The system
is described by  $N=nV$  {\it point} particles at quenched random
positions ${\bf r}_i$, contained in a volume $V=L^d$. Each DPD
particle is a mesoscopic subsystem, that interacts with all
particles that are inside its interaction sphere of radius $r_c$.
The only dynamical variable is the internal energy of the
particle, $\epsilon_i$, which captures the internal degrees of
freedom of the mesoscopic particle. Its evolution equation is the
Langevin equation \cite{ripoll98,landau},
\be \label{langevin}
d\epsilon_i = {\sum_j}^\prime \lambda(ij) \left( {T_j} - {T_i}
\right) dt + {\sum_j}^\prime a(ij) d{W}_{ij} (t),
\ee
where the prime
indicates the constraint $j \neq i$. The first term on the r.h.s.
is dissipative and specifies that a temperature difference causes
flow of energy. The second term represents the Langevin force,
described as Gaussian white noise,
\be
\label{F-noise} \widetilde{F}_{ij}(t) dt = a(ij) dW_{ij}(t).
\ee
It takes into account thermally induced fluctuations in each
particle causing random exchange of energy between particles.
Conservative forces are absent in this model.  The relaxation
coefficient $\lambda(ij)$ models thermal conduction and $a(ij)$ is
the amplitude of the noise. These model parameters are assumed to
be {\it symmetric} under particle interchange. In principle
$\lambda(ij)$ and $a(ij)$ depend on the relative distance
$r_{ij}$, and on the internal energy of the particles $i$ and $j$.
If $a(ij)$ depends on $\epsilon_i$ and $\epsilon_j$, then
\Ref{F-noise} represents multiplicative noise, because the
internal energies themselves depend also on the noise.
 The
model parameters are of the general form,
\be \label{w-range}
\lambda(ij) = \lambda_{ij} w(r_{ij}); \qquad a (ij) =a_{ij}
w_s(r_{ij}).
\ee
The {\it range} or interaction functions $w(r)$ are $w_s(r)$ are
positive, they vanish for $r>r_c$ and have a finite non-vanishing
value at the origin. Moreover, we choose the following normalization
for $w(r)$, \be \label{norm-w} [w] \equiv \int d\br w(r) =r^d_c \ee
and we will see below that the relation $w_s(r) = \sqrt{w(r)}$ is
imposed by the detailed balance conditions.

To define the temperature $T_i$ of a mesoscopic particle with
energy $\epsilon_i$ the equation of state has to be specified, for
which we use the entropy, or equivalently, the density of internal
states.  The simplest choice is the entropy for an ideal solid,
\be \label{def.s}
s(\epsilon) = C_v \ln (\epsilon/\epsilon_u),
\ee
where $C_v = \alpha k_B$ is the heat capacity of a DPD particle,
which is assumed to be a constant, independent of $\epsilon$. The
parameter $\epsilon_u$ is a constant reference energy, and
represents an additive constant to the entropy. Consequently, the
temperature of a DPD particle follows from $ds/d\epsilon
=1/T(\epsilon)$, and is given by,
\be \label{ecstate}
\epsilon_i = C_v T(\epsilon_i) = \alpha k_B T_i  .
\ee
The dimensionless number $\alpha=C_v/k_B$ is a measure of the {\it
size} of the DPD particle because it scales like the {\it number}
of internal degrees of freedom of the particle. Hence, $\alpha$ is
a {\it large} number, and subleading corrections of relative order
$1/\alpha$ will be consistently neglected in this paper.

The {\it total energy}, $E=\sum_i \epsilon_i $, has only internal
energy contributions, and is exactly {\it conserved} by
construction, because the r.h.s. of (\ref{langevin}), when summed
over $i$, is  chosen to be anti-symmetric under particle
interchange, and therefore vanishes. Consequently, the increments
of the Wiener process associated with the heat conduction have
  to be antisymmetric $d W_{ij}= - d W_{ji}$. The noise term
represents Gaussian white noise with a mean $\overline{dW}=0$ and
with a noise strength,
\be \label{ito.e}
 \overline{d W_{ij}(t) d W_{i^\prime j^\prime}(s)} = (\delta_{ii^\prime} \delta_{j
j^\prime} - \delta_{ i j^\prime} \delta_{j i^\prime})\;
\mbox{min}\{dt, ds\},
\ee
where $\overline{dWdW}$ represents an average over the random
noise, and min$\{a,b\}$ denotes the minimum of $a$ and $b$.
The corresponding Fokker Planck equation for the time evolution of the
$N$ particle distribution, $ \rho({\bf X},t)$ in the phase space
given by ${\bf X} = \{ {\bf x}_i = ({\bf r}_i, \epsilon_i)| i=1,2
\cdots N \}$, reads
\be \label{Liouville}
\partial_t \rho =L\rho.
\ee
If the stochastic differential is interpreted according to \^Ito
\cite{mort69,Gardiner,horst-book,broeck-parr97} the Fokker Planck or
Liouville operator $L$ and its adjoint $L^\dagger$ have to be
written as,
\ba \label{L+dagger}
L &=&  \sum_{i < j}\partial_{ij} \left[\lambda(ij) \left( {T_i} -
{T_j} \right) + \frac{1}{2}  \partial_{ij}  a^2(ij)\right]
\nn L^\dagger &=& \sum_{i < j} \left[\lambda(ij)
\left( {T_j} - {T_i} \right) + \frac{1}{2}
 a^2(ij) \partial_{ij} \right]\partial_{ij},
\ea
where $\partial_{ij}=\partial/\partial \epsilon_i
-\partial/\partial \epsilon_j$. Consequently the Fokker Planck
operator has the standard form, appropriate for a multi-particle
Fokker Planck equation. In Appendix A we construct an ${\cal H}$
function and prove an ${\cal H}$ theorem, i.e. $\partial_t{\cal H}
\leq 0$, which holds provided the dissipation coefficient $\lambda
(ij)$ and the noise strength $a (ij)$ satisfy the so-called {\it
detailed balance } conditions,
\be \label{DB}
a^2(ij) = 2 k_B \lambda (ij) T_iT_j \quad \mbox{and} \quad
\partial_{ij} a^2(ij) =0 \qquad (\forall \{i,j\}),
\ee
which implies that $w^2_s(r) = w(r)$, as derived in Appendix A.
Inserting this relation into \Ref{langevin} puts the Langevin
equation into the form,
\be \label{DB-langevin}
 dT_i = \frac{1}{\alpha k_B} d\epsilon_i= \frac{1}{\alpha k_B}
{\sum_j}^\prime \left[ \lambda(ij) (T_j-T_i)dt +
\sqrt{2 k_B \lambda(ij) T_iT_j} \; dW_{ij} \right],
\ee
where the fluctuating term represents multiplicative noise. To
further specify $\lambda_{ij}$ we consider a subsystem 1 (here a
mesoscopic particle) with energy $\epsilon_1 =C_vT_1$, in contact
with a second subsystem of temperature $T_2$. Then irreversible
thermodynamics gives for the energy relaxation in subsystem 1,
\be \label{irr-thermo}
\frac{dT_1}{dt} = \left(\frac{1}{C_v}\right)\frac{d\epsilon_1}{dt}
= \frac{\lambda_{12}}{\alpha k_B}(T_2-T_1),
\ee
where the specific heat of the system is $C_v =\alpha k_B$, and
the relaxation coefficient $\lambda_{12}= \alpha k_B \lambda_0 $
is a material constant, independent of the energies
$\{\epsilon_i,\epsilon_j\}$ of the interacting subsystems and
proportional to the size $\alpha$ of subsystem 1. So, we choose
\be
\label{lambda-const} \lambda_{ij} = \alpha k_B \lambda_0,
\ee
which defines the relaxation parameter in the Langevin equation
for our heat conduction model, which reads finally,
\be \label{L-fin}
 dT_i = {\sum_j}^\prime \left[ \lambda_0 w(r_{ij})(T_j-T_i) dt + \sqrt{ 2
\alpha^{-1} \lambda_0 w(r_{ij}) T_iT_j} \; dW_{ij} \right],
\ee
where $T_i,T_j$ may be replaced  by  $ \epsilon_i,\epsilon_j$.
When performing simulations it is convenient to make the variables
in the Langevin equation dimensionless, i.e. we express distances
as $\widetilde{r}=r/r_c $, the time  as $\widetilde{t} =\lambda_0
t$, the increment of the Wiener process as $d\widetilde{W}_{ij} =
\sqrt{\lambda_0} dW_{ij}$, and the temperature as $\widetilde{T}_i
= T_i/T_u$, where $T_u$ is an arbitrary reference temperature.  It
follows  by setting $\lambda_0=1$ in the equation above. Here
$d\widetilde{W}_{ij}$ satisfies the relation \Ref{ito.e} with $dt$
 and $ds$ replaced by $d \tilde{t}$ and $d \tilde{s}$

The corresponding Fokker Planck operators take the form,
\ba \label{L-DB}
L &=& \alpha k_B \lambda_0\sum_{i < j} w(r_{ij})\partial_{ij}[
{T_i} - {T_j} +k_B \partial_{ij} T_iT_j]
\nn L^\dagger &=& \alpha k_B \lambda_0\sum_{i < j} w(r_{ij})
[ {T_j} - {T_i} +k_B T_iT_j  \partial_{ij}]
\partial_{ij}.
\ea
From the discussion in Appendix A leading to \Ref{commute} it
follows that $a^2(ij)$ and $\partial_{ij}$ in \Ref{L+dagger}, and
$T_i T_j$ and $\partial_{ij}$ in \Ref{L-DB} are {\it commuting}
to leading order in $1/\alpha$.
This implies that the \^Ito and Stratonovich interpretations are
coinciding, and  that the stochastic differential $dW_{ij }$ can
be treated as a differential in ordinary differential calculus.

In the present context the Fokker-Planck equation is frequently
referred to as Liouville equation. In the same spirit the
corresponding mesoscopic  Langevin equations are referred to as
microscopic equations. To complete the microscopic description we
derive the local conservation law for the microscopic energy
density. This supplies us with a microscopic expression for the
energy flux. It will be used in the  simulations to measure the
macroscopic heat current and the heat conductivity.

 We first introduce the static (quenched) particle density and
 the dynamic energy density, given respectively by,
\be \label{n+e}
\widehat{n}({\bf r}) = \sum_i\delta({\bf r}-{\bf r}_i); \qquad
\widehat{e}({\bf r}) = \sum_i \epsilon_i \delta({\bf r}-{\bf r}_i).
\ee
Following standard arguments we derive an expression for the
local {\em microscopic} energy flux $\widehat{q}({\bf r})$ as well
as for the total flux $\widehat{Q}= \int d{\bf r}
\widehat{q}({\bf r})$. A hat on a symbol denotes a mesoscopic
quantity.

The equation of motion for the expectation value $\langle
\widehat{e}({\bf r}) \rangle_t=e({\bf r},t)$
becomes,
\be \label{efhc.ev}
\partial_t e({\bf r},t) =
\int d{\bf X} \widehat{e}({\bf r} | {\bf X}) \partial_t \rho \equiv
\langle L^\dagger \widehat{e}({\bf r}) \rangle_t = - \bnabla \cdot
\langle \widehat{\bf q}({\bf r}) \rangle_t.
\ee
The last equality in (\ref{efhc.ev}) shows the local energy
conservation law. We have further used the relation,
\ba \label{L+er}
&L^\dagger \widehat{e}({\bf r}) =  \sum_{i < j} w(r_{ij}) \lambda_{ij}
({T_j} - {T_i}) \left[ \delta({\bf r} - {\bf r}_i) - \delta({\bf
r} - {\bf r}_j) \right]&
\nn
&\simeq - \bnabla \cdot \sum_{i < j} w(r_{ij}) {\bf r}_{ij}
\lambda_{ij} ({T_j} -{T_i})\delta({\bf r} - {\bf r}_i) \equiv -
\bnabla \cdot \widehat{\bf q}({\bf r})&,
\ea
and note that the terms in the Fokker Planck operator \Ref{L-DB},
coming from  the noise,  do not contribute to the macroscopic
flux of energy. The last line has been obtained by expanding
$\delta({\bf r} - {\bf r}_j)$ in powers of ${\bf r}_{ij}$,  i.e.
$\delta({\bf r} - {\bf r}_i + {\bf r}_{ij}) \simeq \delta({\bf r} -
{\bf r}_i) + {\bf r}_{ij} \cdot \bnabla \delta({\bf r} - {\bf r}_i)
+ {\cal O} ( \nabla^2)$. The  total microscopic heat flux,
$\widehat{\bf Q} \equiv  \int d{\bf r} \widehat{{\bf q}}({\bf r})=
\widehat{\bf Q}_D + \widehat{\bf Q}_R$, consists of a dissipative
(D) and a fluctuating (R) part \cite{ernst05}. With the help of
the relations $ \lambda_{ij}=
\alpha k_B\lambda_0$ and $\epsilon_i=\alpha k_B T_i$ the
dissipative part is given by,
\be \label{qtotal}
\widehat{{\bf Q}}_D=
\lambda_0 \sum_{i < j} w(r_{ij}) {\bf r}_{ij} ({\epsilon_j} -
{\epsilon_i}),
\ee
where terms of order $1/\alpha$ have been neglected. We note that
the current $\widehat{\bf Q}_D$ does not contain kinetic
contributions, but is a sum of pair contributions involving the
dissipative interactions. This is the mechanism of collisional
transfer, representing instantaneous transfer of energy to particle
$i$ from all particles $j$ in the interaction sphere defined by
$r_{ij} \leq r_c$. The current $\widehat{\bf Q}_D$ should be
compared with the contributions to the microscopic stress tensor
involving the conservative interaction potentials in Hamiltonian
fluids. Furthermore, the expression for ${\bf \widehat{Q}}_D$
also illustrates that the total macroscopic heat current is
determined by the energy difference, $\epsilon_j -\epsilon_i$,
 i.e. by the "temperature gradient" between $\br_i$ and $\br_j$.
The fluctuating part $\widehat{\bf Q}_R = -\sum_{i<j} a(ij) {\bf
r}_{ij} \widetilde{F}(ij)$ with $\widetilde{F}_{ij}$ in Eq.
\Ref{F-noise} and $\av{\widehat{Q}_R}_t=0$.

The macroscopic energy flux ${\bf q}$ obeys the standard linear
constitutive law of irreversible thermodynamics,
\be \label{q.macr}
{\bf q} = \langle \widehat{\bf q} \rangle_t = - \lambda \bnabla T ,
\ee
where $\lambda$ is the coefficient of the heat conductivity.
Combination of (\ref{efhc.ev}) and (\ref{q.macr}) with the
relation $\delta e = n C_v \delta T  = n\alpha  k_B \delta T$,
yields the equation for heat diffusion,
\be \label{k.Dt}
\partial_t T = \frac{\lambda}{n \alpha k_B} \nabla^2 T
\equiv D_T \nabla^2 T,
\ee
where $D_T$ is the heat diffusivity.

For later reference we mention that the heat conductivity can also
be calculated and/or simulated using the {\it equilibrium time
correlation} function of $\widehat{\bf Q} = \widehat{\bf Q}_D +
\widehat{\bf Q}_R$. As the energy density, $e(\br,t) \equiv \alpha k_B n(\br)
T(\br,t)$, satisfies a local conservation equation, the heat
conductivity can be expressed in the Green-Kubo formula,
\be \label{kubo}
\lambda = \frac{1}{d V k_B T^2} \int_0^\infty dt \av{{\bf
\widehat{Q}(t) \cdot \widehat{Q}(0)}}_{0},
\ee
where the average $\av{\cdots}_0$ is taken over the apropriate
equilibrium ensemble. On the basis of the analogy between the
Liouville  and the Fokker Planck operators in \Ref{L-DB}, with
$T_i$ replaced by $\epsilon_i/\alpha k_B $, any of the standard
derivations of these formulas
\cite{hansen-mcdonald,Dorfman-ernst-JSP} carries over directly to
our DPD solid. We further note that the microscopic energy current
${\bf \widehat{Q}(t)}$ does not contain any "subtracted part"
because this model does not have any conserved quantity with a
vector character, such as the total momentum.

For the case of general DPD fluids the Green-Kubo formula for the
viscosity has been derived in Ref. \cite{pep95-kubo}. Another
representation of the transport coefficients, equivalent to the
Green-Kubo formulas, is given by the so-called Helfand formulas
\cite{helfand,Allan+T-book}. It reads for the present case,
\be \label{helfand}
\lambda = \lim_{t \to \infty}\frac{1}{2 Vk_B T^2} \frac{d}{dt}
\av{(M(t)-M(0))^2}_0
\ee
with $M$ given by,
\be \label{moment}
M(t) = \sum_i \epsilon_i(t) x_i ,
\ee
One easily verifies that the microscopic heat current, $Q_x =
L^\dagger M$, in \Ref{kubo} can be obtained from $M$. The Helfand
formulas are generalizations of Einstein's formula for the
coefficient of self diffusion, and are presumably more convenient
in numerical simulations than the Green-Kubo formulas.

A further consequence of the ${\cal H}$ theorem, discussed in
Appendix A, is the existence of a unique equilibrium state, the
Gibbs' state. Its explicit form has also been determined in
Appendix A. In the main text of this paper we only need the single
particle equilibrium distribution function for the DPD solid, as
derived in \Ref{a11}, i.e.
\begin{equation} \label{fcan}
\psi_0(\epsilon) =
\frac{\beta}{\Gamma(\alpha+1)}\left(\beta\epsilon\right)^\alpha
\exp[-\beta\epsilon],
\end{equation}
where $\alpha$ is a measure for the number of internal degrees of
freedom. In Ref. \cite{ripoll98,bonet2,Ripoll-thesis} simulations
of the equilibrium distributions in the conduction model have been
performed, and good agreement between the simulation results and
the analytical expressions has been obtained. For instance, the
simulated energy fluctuations  agree very well with the
theoretical prediction $\av{(\delta\epsilon_i)^2} =
(\alpha+1)/\beta^2 \simeq \alpha/\beta^2$ within error bars
smaller the $0.7\%$ \cite{Ripoll-thesis}.

\renewcommand{\theequation}{III.\arabic{equation}}
\setcounter{section}{2} \setcounter{equation}{0}
\section{Heat conductivity}
\subsection{Mean field theory}

The DPD model for heat conduction is expected to produce a
macroscopic behavior described by a macroscopic heat equation. Our
aim is to prove this assertion and to relate the effective thermal
diffusivity appearing in the macroscopic heat equation to the model
parameter $\lambda_0$ and the range function $w(r)$. To do so we
will use a mean field approximation. We start with an a priori
estimate of the transport coefficient.

In a naive kinetic picture of the relevant transport mechanism,
used in Ref.\cite{ripoll98}, amounts of heat or energy hop on a
random lattice with an average lattice distance $l_s = n^{-1/d}$,
and a hopping frequency $\omega_0$. This picture, based on
{kinetic} transport of energy, leads to a heat diffusivity $D_0=
\omega_0 l^2_s$, where $\omega_0$ is the decay rate of an energy
or temperature fluctuation. As the decay rate $\omega_0 \propto n$
(see below), this would lead to an a priori estimate $D_0 \propto
n^{1-d/2}$, which does not hold for the collisional transfer
mechanism.

As the DPD particles are quenched, there is {\it no kinetic}
transport, but only {\it collisional transfer} of energy, i.e.
{\it instantaneous} transfer of energy through particle
interactions. It takes place only over distances less than the
range $r_c$ of the interactions. This picture leads to a
diffusivity on the order of $D = \omega_0 r^2_c$, where $r_c$ is
the range of the interaction function $w(R)$, and $\omega_0$ is a
typical frequency. For {\it large} densities this frequency can be
estimated from the first term on the r.h.s. of \Ref{L-fin} as,
\be \label{4.2}
\omega_0 = \lambda_0 {\sum_j}^\prime \av{w(r_{ij})} \simeq
\lambda_0 n [w] = \rho \lambda_0 ,
\ee
where $[w]$ is defined in \Ref{norm-w} and $\av{\cdots}$ denotes
an average over all quenched particles. Here the {\it reduced}
density $\rho \equiv nr_c^d$ is on the order of the mean number of
interacting neighbors with $r_{ij} \leq r_c$, surrounding the
$i-$th particle.  The freedom to shift constant factors in
\Ref{w-range} from $\lambda_0$ to $w(r)$ is fixed by the
normalization $[w] = r^d_c$ in \Ref{norm-w}. The {\it a priori
estimate} for the diffusivity at large densities is then,
\be \label{4.3}
D \simeq \omega_0 r^2_c =  \rho\lambda_0 r_c^2 .
\ee
The estimate \Ref{4.3} will be confirmed by detailed kinetic
theory calculations.

To calculate the heat flux we express the average of \Ref{qtotal}
in terms of the two-particle distribution function, $f^{(2)}(
\br_1,\epsilon_1,\br_2,\epsilon_2 ,t)$, yielding
\be  \label{4.4}
{\bf q} =  \frac{1}{2}\lambda_0 \int d\epsilon_1 d\epsilon_2 \int
d {\bf R} w(R) {\bf R} (\epsilon_2 -\epsilon_1 )
 f^{(2)} ({\bf r}, \epsilon_1, {\bf r} -{\bf R}, \epsilon_2, t).
\ee
The basic ansatz in this {\it mean field} theory is that the fluid
rapidly relaxes to a {\em local equilibrium} described by the
local fields $b(\br,t)$, being the temperature $T({\bf r}, t) = 1/
k_B \beta({\bf r}, t)$ and local (quenched) density $n({\bf r})$.
This happens on a time scale $1/\omega_0$ where $\omega_0 \sim
\lambda_0\rho (\br)$ is the local relaxation rate of the
temperature fluctuations and $\rho (\br)$ is on the order of the
number of particles inside the sphere centered at $\br$.
Temperature gradients, which are smoothly varying in space, only
build up on a hydrodynamic time scale, as described by Fourier's
heat law.
  Because {\it  conservative} forces are absent in our model, the
 {\it local equilibrium} pair distribution function is {\it
exactly} equal to the corresponding pair function of an ideal gas,
i.e. it is simply a product of single particle local equilibrium
distribution functions $f_0$, {\em i.e.}
\be\label{4.5}
f^{(2)} ({\bf r}, \epsilon, {\bf r}^{\prime}, \epsilon^{\prime},
t) =  f_0 (\epsilon| b({\bf r}, t) ) f_0(\epsilon^\prime|b({\bf
r}^{\prime}, t)),
\ee
which depends on $b({\bf r}, t)$ and $b({\bf r^\prime}, t)$. Their
explicit form follows from (\ref{fcan}) as,
\be \label{4.6}
f_0(\epsilon|b({\bf r})) = n(\br) \psi_0(\epsilon|\beta (\br)) =
n(\br)
\beta({\br}) \left[ \beta({\br}) \epsilon \right]^\alpha
e^{-\beta({\br}) \epsilon} / \Gamma(\alpha+1),
\ee
where $\psi_0$ only depends on the  temperature at the position
${\bf r}$ of the particle.

To derive an expression for the heat conductivity in \Ref{q.macr}
the equations \Ref{4.4}-\Ref{4.6} need to be expanded in
gradients, i.e. $\beta ({\bf r -R}) = \beta - {\bf R} \cdot
\bnabla \beta +{\cal O}(\nabla^2)$, and similarly for $ n ({\bf r
-R})$. Because the integrand in \Ref{4.4} is {\it anti-symmetric}
in both ${\bf R}$ and $(\epsilon_2-\epsilon_1)$, the only
non-vanishing contribution to ${\bf q}$ from $f^{(2)}$ in
\Ref{4.5} must be linear in ${\bf R}$ and $\epsilon_2$. The result
is,
\be \label{4.7}
f_0(\epsilon_2| \beta({\bf r -R})) \to - n ({\bf R} \cdot
{\bnabla}
\beta ) \psi_0 (\epsilon_2) \frac{\partial}{\partial \beta}
\ln \psi_0(\epsilon_2) \to n \epsilon_2 ({\bf R} \cdot {\bnabla}
\beta) \psi_0(\epsilon_2).
\ee
Symmetrizing the last expression, $\epsilon_2 \to  \half(
\epsilon_2 - \epsilon_1)$,  and inserting this in \Ref{4.5} and
\Ref{4.6} yields after some algebra for the heat current,
\be \label{4.8}
{\bf q} = - \textstyle{\frac{1}{4}} \lambda_0 k_B n^2 \int d {\bf
R} w(R){ \bf R}{ \bf R} \cdot  \bnabla T \;\beta^2 \av{(\epsilon_2
-\epsilon_1)^2}_0 \equiv -\lambda {\bnabla} T,
\ee
yielding a mean field prediction for the {\it heat conductivity},
\be \label{4.9}
\lambda (\rho) = \frac{\alpha}{2d} \lambda_0 k_B n^2 [w]
\av{R^2}_w \equiv \lambda_\infty (\rho),
\ee
which is quadratic in the density. We refer to this mean field
value \Ref{4.9} as the {\it saturation} or high density value.
This value is expected to be exact when the difference between the
actual number of particles inside an interaction sphere and its
mean value can be neglected. Here $ \av{(\epsilon_2
-\epsilon_1)^2} \simeq 2\alpha/\beta^2$ denotes an average over
the canonical ensemble \Ref{rofact}, and is given by
\Ref{pe.moments}. Moreover the following relation has been used,
\be \label{4.10}
\int d{\bf R} R_\alpha R_\beta w({R}) = \frac{1}{d}\delta_{\alpha
\beta} \int d {\bf R} R^2 w({ R}) \equiv \frac{1}{d}
\delta_{\alpha \beta}[w] \langle R^2 \rangle_w ,
\ee
where the last equality defines the normalized second moment
$\langle R^2 \rangle_w$. The mean field value for the {\it heat
diffusivity}, defined in \Ref{k.Dt}, is obtained similarly,
\be \label{4.11}
D_\infty = \frac{\lambda_\infty}{n\alpha k_B} =   \frac{\omega_0
}{2d}\av{R^2}_w,
\ee
where $\omega_0 =\lambda_0 n [w] =\lambda_0 \rho$ is the typical
decay rate of an energy fluctuation.

In the literature on DPD simulations different choices of the
range function $w(r)$ have been considered, for instance,
\be \label{w-lucy}
w(r) = \left\{ \begin{array}{lll}
 A_d \theta (r_c -r) &\qquad &
\mbox{(Heaviside)}\\
 B_d \left(1-\frac{r}{r_c}\right)\theta (r_c -r)&\qquad
 &\mbox{(Triangle)} \\
C_d \left(1+3\frac{r}{r_c}\right) \left(1-\frac{r}{r_c}\right)^3
\theta (r_c -r)& \qquad &\mbox{(Lucy)}
\end{array} \right.
\ee
Here $\theta(x)$ is the Heaviside or unit step function, and the
constants $A_d, B_d, C_d $ are normalized such that $[w]= r_c^d$
in $d-$dimensions. Calculation of the moments $\langle R^2
\rangle_w$  yields then,
 \be
\label{D-lucy} D_\infty= \frac{\lambda_\infty}{n \alpha k_B} =
\left\{
\begin{array}{lll}
\frac{1}{2(d+2)}  \omega_0
r_c^2 & \qquad & \mbox{(Heaviside)}\\[2mm]
\frac{d+1}{2(d+2)(d+3)}
\omega_0 r_c^2 & \qquad & \mbox{(Triangle)}\\[2mm]
\frac{d+3}{2(d+5)(d+6)}\omega_0 r_c^2 & \qquad & \mbox{(Lucy)}
\end{array}
\right.
\ee

Before concluding this subsection we make a number of comments. If
we would have chosen a normalization of $w(r)$, different from
\Ref{norm-w}, say $w'(r) = {\cal C} w(r)$, while keeping the unit
of time  fixed, the diffusivity would change to $D' = {\cal C}D$.
We recall that the heat current in \Ref{4.4}, which is based on
the mechanism of collisional transfer, establishes itself {\it
instantaneously}, i.e. a high frequency limit occurring  on the
fast time scale $1/\omega_0$, because pair interactions transfer
energy instantaneously between $\br_i$ and $\br_j$, whenever
$r_{ij} \leq r_c$. Such heat currents only occur in {\it dense}
systems, and they are non-vanishing in a state of local
equilibrium. On the other hand kinetic transport of energy,
carried by moving particles, establishes itself only on the slower
hydrodynamic time scale.\\
We add a technical remark. In general the heat flux in \Ref{4.4}
would also pick up contributions from the additional gradient
terms in the single particle distribution functions in \Ref{4.5},
\be \label{4.10a}
f(x_i)= f_0(x_i)[ 1+H(\epsilon_i|\beta(\br_i))\nabla\beta(\br_i)+
\cdots] .
\ee
with $i=1,2$, that would have to be added to \Ref{4.5}.  However,
such terms contribute to the heat flux ${\bf q}$ only terms of
${\cal O}( (\nabla T)^2)$ and ${\cal O} (\nabla^2 T ) $ because of
the parity in ${\bf R}$ of the integrand in \Ref{4.4}.\\
Next we note that the results \Ref{4.11} and \Ref{D-lucy} above are
in qualitative agreement with the a priori estimate, $D \simeq
\omega_0 r^2_c$ in (\ref{4.3}). The expressions for $D_T$ in
\Ref{4.11} and \Ref{D-lucy}, with physical dimensions $L^2/t$, has
the typical structure of a diffusivity, i.e. a collision frequency
multiplied with the square of the interaction range $r_c$, over
which energy is transported by the mechanism of collisional
transfer.

Finally we note that the heat diffusivity and the typical
frequency $\omega_0 $ are proportional to the reduced density, $
\rho=n r_c^d$, whereas the heat conductivity $\lambda_\infty$ in
\Ref{4.9} is proportional to ${\rho}^2$. In the DPD fluid with
viscous dissipation, studied in Ref. \cite{mbe1}, there are of
course {\it kinetic} contributions to the viscosity as well, apart
from the collisional transfer (ct)
 contribution to the kinematic viscosity, $\nu_{ct} \sim \rho$,
 being a diffusivity, and to the shear viscosity, $\eta_{ct}
\sim \rho \nu_{ct} \sim {\rho}^2$.  Also in the Enskog theory for
a dense fluid  of hard spheres \cite{chapman} similar collisional
transfer contributions to the transport coefficients are present,
 sometimes referred to as the {\it instantaneous} transport
 coefficients, $\eta_\infty, \lambda_\infty$,
for the reasons explained above. The ${\rho}^2-$ density
dependence of the heat conductivity  in \Ref{4.9} is a direct
consequence of the collisional transfer mechanism. For
sufficiently {\it high} densities, the heat flux ${\bf q }$ is
proportional to the local density of interacting pairs, $\sim
{\rho}^2$, and the heat conductivity is given by its saturation
value $\lambda_\infty (\rho)$ in \Ref{4.9}. This prediction holds
when the typical density fluctuations $\delta \rho \sim
\sqrt{\rho} $ can be neglected with respect to the mean value
$\rho$.

\subsection{Simulations and conductivity threshold}

In our simulations we use the Langevin equation \Ref{L-fin}   to
determine the  dynamic properties of the system. The physical
system is assumed to be a $d$-dimensional cubic box of volume
$V=L^d$, with a cold and a hot wall at  $x=0$ and $x=L$
respectively. This is achieved by putting two extra layers of
particles at each boundary, filled with particles at a constant
temperature.  The layers have a width $r_c$ to ensure that any
particle inside the system interacts on average with the same
number of neighboring particles. Therefore, these extra layers act
as {\em thermal baths} which are prepared at temperatures $T_{\rm
cold}$ and $T_{\rm hot}$ such that a gradient $(T_{\rm hot}-T_{\rm
cold})/L$ is applied to the system. In the remaining directions we
impose periodic boundary conditions. Initially the box is seeded
with $N$ mesoscopic (point) particles located at random,
surrounded by overlapping interaction spheres of radius $r_c$.
The initial temperature of the particles is $(T_{\rm hot}-T_{\rm cold})/2$.
Another relevant quantity is the {\it mean number of interacting
neighbors}, $\nu (r_c) =\av{\widehat{\nu}_i}$ inside the sphere
$r_{ij} \leq r_c$, where
\ba \label{nn-number}
 \widehat{\nu}_i&=& \widehat{\rho}_i\varpi_d  \equiv {\sum_j}^\prime\theta (r_c - r_{ij})
 \nn \nu(r_c) &\simeq& n  \varpi_d r_c^d = \rho \varpi_d
\quad (\mbox{large}\;\; \rho).
\ea
Here $\rho$ is the reduced density, and $\varpi_d = \pi^{d/2}
/\Gamma (1+d/2)$ is the volume of  a $d-$dimensional unit sphere
$(d=1,2,\cdots)$. The fluctuating variables $\widehat{\nu}_i=
\varpi_d \widehat{\rho}_i$ are subject to large static
fluctuations, especially at {\it small} densities, and are
distributed according to a Poisson distribution.

We perform a series of simulations by numerically integrating the
equation (\ref{L-fin}) with a given temperature gradient and
compute the macroscopic heat flux in the steady state.  The time
required to reach the steady state increases with decreasing
density, and even {\it diverges} as the density approaches a {\it
threshold} value, to be discussed below. The heat flux is
calculated with the help of definition (\ref{qtotal}). This
expression involves a large number of pairs of particles which
guarantees reasonably good statistics. If the density is not too
close to the threshold, the simulations show a linear temperature
profile between the two heat baths, as predicted by Fourier's law.
The measured heat flux is found to be linear in the applied
temperature gradient, from which the heat conductivity can be
extracted. Therefore, since this linear relation between $Q_x$ and
$\nabla_x T$ has been confirmed by the simulations, a single
measurement of the macroscopic heat flux for a given gradient
suffices to obtain the thermal diffusivity. Statistical errors can
be estimated by making an average over several independent runs.
In principle, alternative ways to measure the heat conductivity
would be to simulate the Green-Kubo formula \Ref{kubo}, or to set
up a sinusoidal intitial temperature field $T_i(0)$, and to
measure $D_T$ from the decay rate of the initial temperatures
\cite{gerits}.

For our further discussions it is convenient to introduce,
\be \label{ratio}
R_d(\rho)= D_T(\rho)/D_\infty =\lambda(\rho)/\lambda_\infty
(\rho).
\ee
We first consider the transport properties in the
three-dimensional heat conduction model, and use the simulation
results, obtained in Ref. \cite{ripoll98,Ripoll-thesis}, where the
range function $w(r)$ was chosen to be the Lucy function
\Ref{w-lucy}, and we compare the measurements with the analytic
predictions \Ref{D-lucy}, as shown in Fig.1. The simulation
results were performed with random spatial configurations. They
approach the theoretical predictions at {\it high} densities.
However, as the density decreases, the ratio $R_3$ rapidly
decreases, almost by a factor 2 at the lowest densities simulated
$( \rho \simeq 3.8)$. In these measurements averages over
different values of the parameters $N$ and $\alpha$ have been
taken at fixed reduced density $\rho$. Both $N$ and $\alpha$
should be sufficiently large to reduce the finite size effects and
improve the statistics.

\begin{figure}[h]
\includegraphics[width=5.5cm,angle=-90]{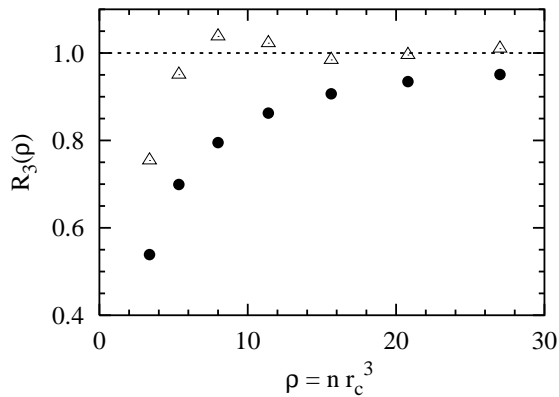}
\caption{Simulated values of the thermal diffusivity $R_3(\rho) =
D_{T}/D_\infty$ versus $\rho$ for the 3-D heat conduction model
with $N= 10^3 $ DPD particles. Results from off-lattice
simulations ($\bullet$) and on-lattice simulations ($\triangle$)
are compared with the mean field value \Ref{D-lucy} (dashed line).
At $\rho =3$ and 27 the system sizes are respectively $L/ r_c =
(N/\rho)^{1/3} \simeq 6.9$ and 3.3, which is rather small, and
strong finite size effects are to be expected.}
\end{figure}

A consistent explanation for these large deviations seems to be
that the mean field theory does not take into account the local
fluctuations in the actual number of particles, $\widehat{
\rho}_i$, in the interaction sphere around particle $\br_i$. These
fluctuations are particularly large at low densities. To test this
working hypothesis we place all $N$ particles on a completely
filled cubic lattice with lattice distance $l_s = (V/N)^{1/3}$,
thus suppressing all local density fluctuations.  The resulting
measurements are represented in Fig.1 by ($\triangle$)'s. Note
that this suppression of density fluctuations considerably extends
the agreement between theory and simulations towards lower
densities. The on-lattice simulations support our working
hypothesis, and the observations are consistent with the good
agreement at high densities, where fluctuations in
${\widehat{\rho}}_i$ are small, but a theoretical explanation of
the $\rho$-dependence of the heat conductivity of our original
random solid is still lacking. The improved agreement between
theory and simulations was here obtained by modifying the model.
Further modifications of the random heat conducting solid model to
suppress the local density fluctuations were introduced in Ref.
\cite{willemsen-JCF00}, but do not increase our understanding of
the density dependence of $D_T(\rho)$ in DPD fluids and solids.

\begin{figure}[h]
\includegraphics[angle=270,width=8cm]{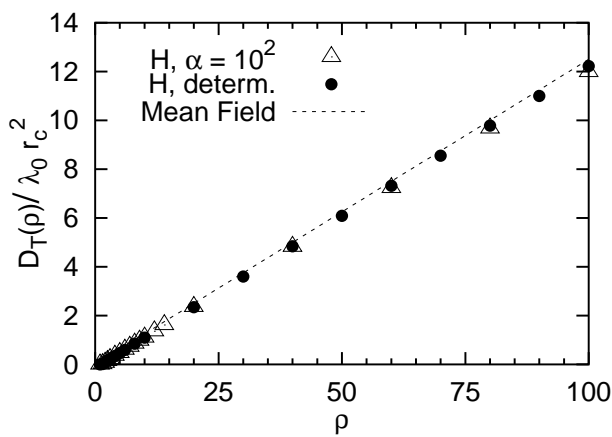}
\caption{2-D simulations of the dimensionless heat diffusivity
$D_T(\rho)/\lambda_0 r_c^2$ plotted versus $\rho$ to test the mean
field prediction $D_\infty(\rho)$ in \Ref{4.9} (dashed line), valid at
high density. Labels ($\triangle$) and ($\bullet$) refer to the random
solid respectively with and without Langevin force in systems with
$10^4$ and $10^5$ DPD particles. H refers to the Heaviside weight
function. }
\end{figure}

To test these concepts the simulation results for the
three-dimensional model in Fig.1 are not very suited, because the
existing 3-dimensional simulations \cite{Ripoll-thesis} suffer
from large finite size effects ($L/r_c=3.3$ or $6.9$). Furthermore
the range function $w(r)$ was taken to be the Lucy function in
\Ref{w-lucy}, which gives larger weights to shorter bonds. To
optimize the similarity with the classic bond percolation
problems, we give all bonds equal weights by taking $w(r)$ as the
Heaviside function. Moreover, to make the simulations less
demanding, we have carried out simulations of our heat conduction
model in two dimensions. Fig.2 shows the good agreement between
the two-dimensional simulations and mean field theory at high
densities, as also observed in the three-dimensional simulations
both of Fig.1, as well as in Fig.1 of Ref. \cite{bonet3}. However,
at lower densities $D_T(\rho)$ decreases faster than linear, as
shown in Fig.3.

To display the connection to percolation it is instructive to plot
$R_2(\rho) \equiv D_T(\rho) /D_\infty(\rho)$, as shown in Fig.3.
The plots strongly suggest the existence of a conductivity
threshold $\rho_c \geq \rho_p$, where $\rho_p = 1.43629 $ is the
threshold value of two-dimensional continuum percolation
\cite{ziff-3d}.

\begin{figure}[hbt]
\includegraphics[angle=270,width=4.9cm]{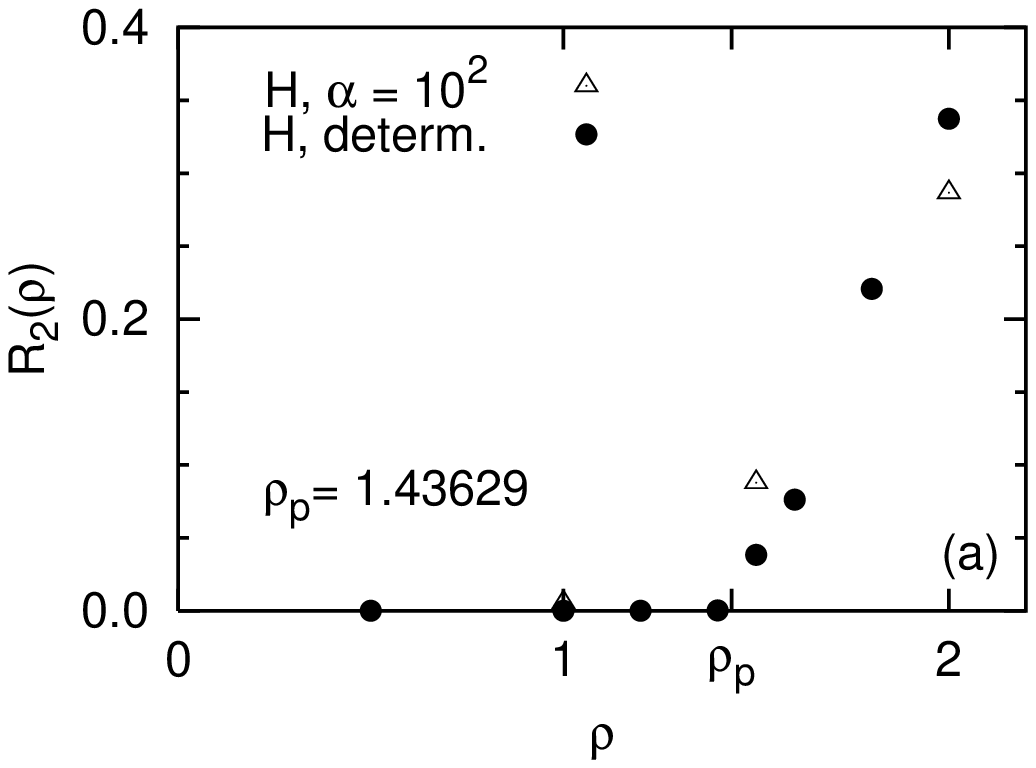}
\includegraphics[angle=270,width=4.9cm]{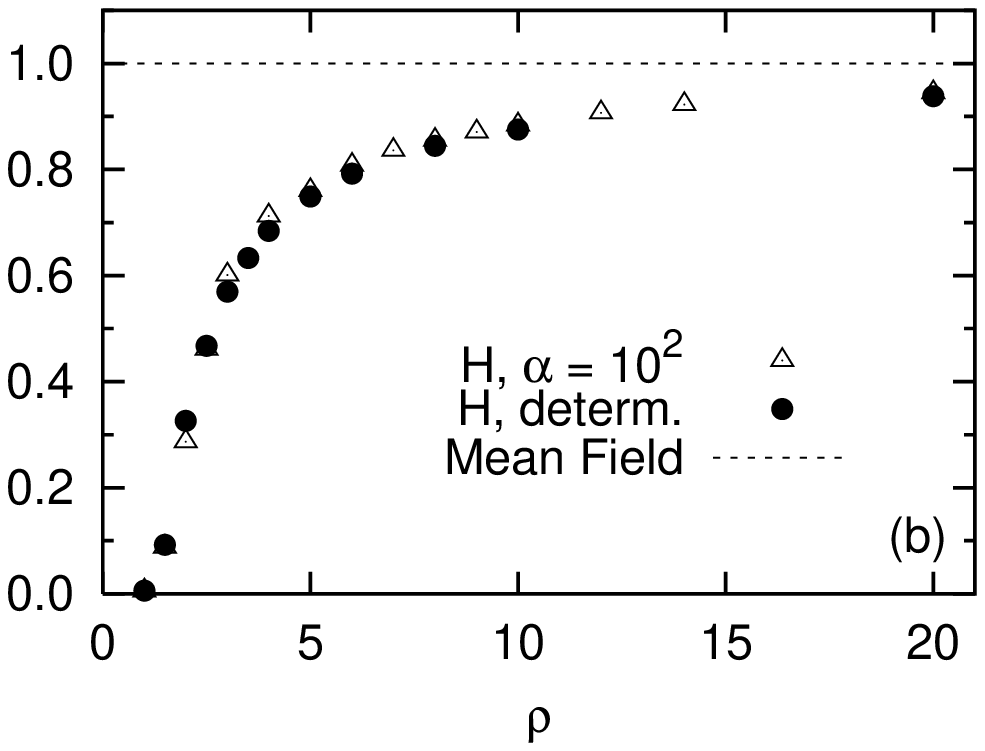}
\includegraphics[angle=270,width=4.9cm]{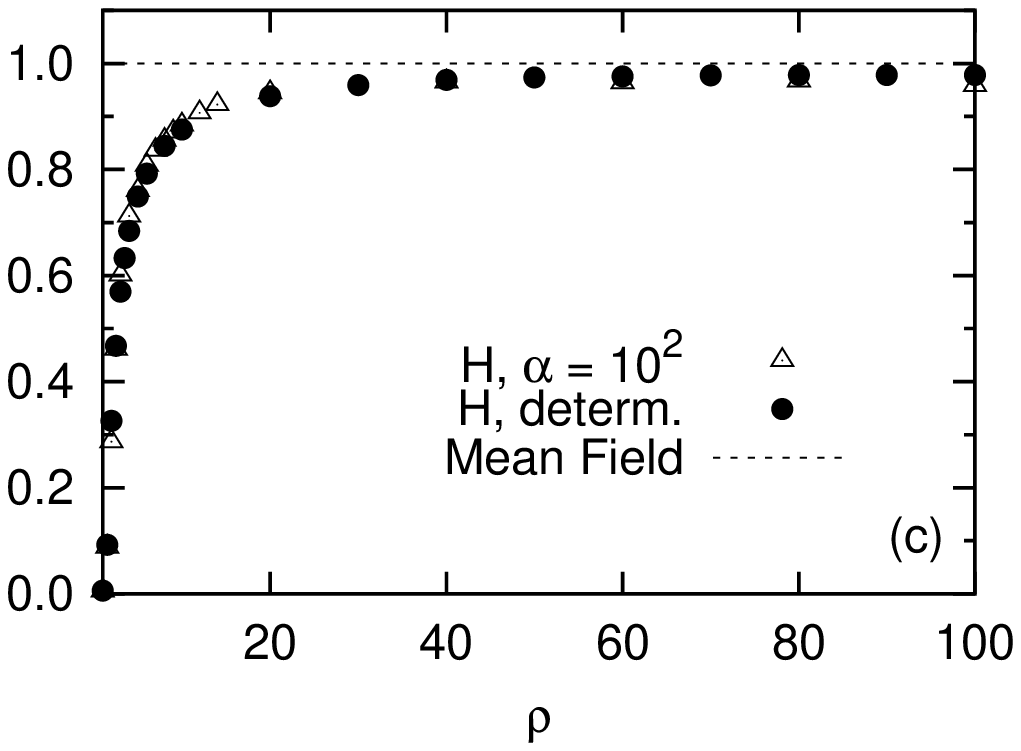}
\caption{$2$-D simulations of the heat diffusivity $R_2(\rho)$
plotted versus $\rho$. Figures a,b,c show respectively the
behavior near threshold, the cross-over, and the approach to the
saturation value at large $\rho$. For definitions of symbols and
parameter values we refer to the previous figure.}
\end{figure}

Fig.3 shows that in our model the conductivity threshold
$1.4<\rho_c<1.5$ which agrees quite well with the known
percolation threshold.  Nevertheless, determining the value of
$\rho_c$ with higher precision becomes more delicate since the
required times for equilibrating the system are diverging as $\rho
\downarrow \rho_c$. Simulations show that the approach to the
expected linear temperature profiles at $\rho \gtrsim 4$ is
relatively fast (relaxation time $t_0 \lesssim 300$). But these
times increase for smaller densities, for $\rho = 1.6 $ ($t_0
\simeq 2.5 \times 10^4$) and for $\rho = 1.4$ ($t_0 \simeq 10^6$).
The profiles at the times of measurement are shown in Fig.4.
Furthermore Figs.3b and 3c show the very slow cross-over of the
conductivity at large $\rho$ to the mean field result, where
density fluctuations are small.

The corresponding threshold in three dimensional continuum
percolation is $\rho_p =0.65296$. This value is not inconsistent
with the low density extrapolation of the three-dimensional heat
diffusivity in Fig.1, but simulation data for the
three-dimensional random solid, sufficiently close to the
conductivity threshold, are lacking in the simulations of Ref.
\cite{ripoll98,Ripoll-thesis}.

 \begin{figure}[hb]
\includegraphics[angle=270,width=8cm]{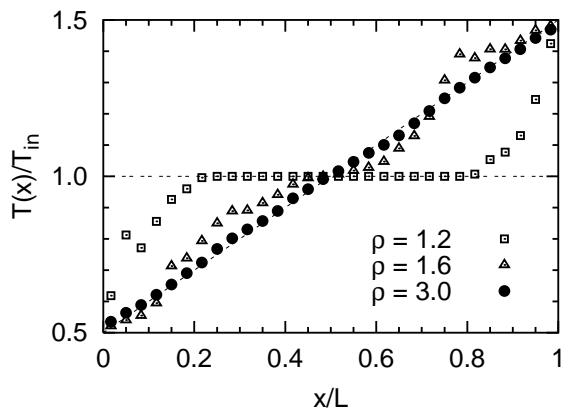}
\caption{Temperature profiles at times of measurement. For $\rho
\gtrsim 3$ a steady linear profile is reached, but this is not the
case for $\rho \lesssim 1.6.$}
\end{figure}

\begin{figure}[h]
\includegraphics[angle=-90,width=7.cm]{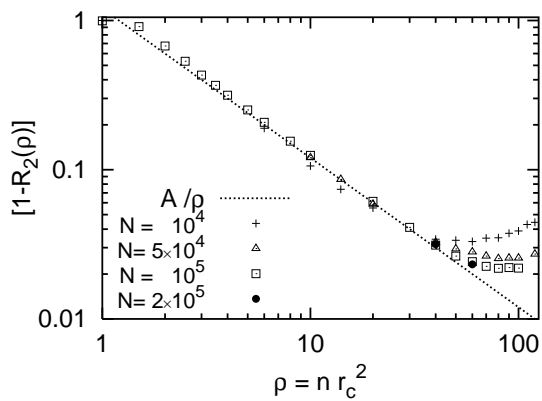}
\caption{The plot shows that the dominant correction to
$1-R_d(\rho)$ behaves in 2D like $A_2/{\rho}$ with $A_2 \simeq
1.2$. The data refer to the deterministic case (vanishing Langevin
force) with the Heaviside weight function.}
\end{figure}

 To further analyze the observed density dependence we show in
 Fig.5 the function $1-R_2(\rho)$ on a log-log plot.
The plot strongly suggests that the subleading correction to the
heat conduction at large $\rho$ has the form, $R_2(\rho) \simeq 1-
A/ \rho$ with $A \simeq 1.2$, but an analytic calculation of the
quantity is lacking. The deviations from the above asymptotic
form, shown at large $\rho$ in Fig.5, are finite size effects. The
simulations on the top curve ($+$), middle curve ($\triangle$) and
bottom curve ($\square$) involve resp $N= 10^4, 5 \times 10^4,
10^5$ particles. If $N$ increases from $10^4$ to $10^5$ at fixed
$\rho$ (say $\rho =50$) the system size increases from $L/r_c =
\sqrt{N/\rho} \simeq 14$ to $44$, and the finite size effects
{\it decrease}. If $N$ and $\rho$ increase by
the same factor, the finite size effects remain of the same order,
and if the density increases from $\rho =50 $ to $\rho=120$ at
fixed $N$ (say $N=10^4$), the system size {\it decreases } from
$L/r_c \simeq 14$ to 9, and the corresponding finite size effects
{\it increase} in Fig.5.  The small three-dimensional
systems seem to be dominated by finite size effects, which appear
already before the subleading term $A/\rho$ becomes dominant.

It is also of interest to illustrate another effect on the heat
diffusivity of the fluctuations. The expected coverage with
'black circles' (see Section I)  or the  black surface fraction at
reduced density $\rho$ is $\varphi (\rho) = 1 - \exp[-\fourth \pi
\rho]$, and the white surface or free volume fraction is
$\exp[-\fourth \pi \rho]$ \cite{ziff-3d}. Then at $\rho =
\{1.43629; 20; 100\}$ there are in each black circle on average $
\fourth \pi \rho -1 = \{0.13; 14.7; 77.5\}$ excess particles, and
the white  free surface fraction, $1- \varphi (\rho) =
\exp[-\fourth \pi \rho] =$
$\{0.32;2\times 10^{-7};8\times 10^{-35}\}$, becomes extremely
small with increasing $\rho$, whereas the corresponding simulated
value of the heat diffusivity is still a sizeable fraction, $1- R_2
(\rho) = \{100\%;10\% ; 3\% \}$, below its saturation value.

\renewcommand{\theequation}{IV.\arabic{equation}}
\setcounter{section}{3} \setcounter{equation}{0}
\section{Generalized Hydrodynamics}

In this section we further explore the analogy between fluids
and statistically disordered solids, using kinetic theory.
Generalized hydrodynamics describes the decay of small spatial
fluctuations in the conserved local densities at different
wavelength, $\lambda_k = 2\pi /k$.  Their decay rates depend
strongly on how the probing wavelength (size of colloidal
particles, polymers, pores....) compares to the range of the DPD
forces. These decay rates can be expressed in terms of
$k-$dependent transport coefficients. The wavelengths cover both
the standard hydrodynamic regime as well as the mesoscopic regime.

A classical method \cite{resibois} to analyze the full
hydrodynamic regime is to determine the eigenvalues (decay rates)
of the Fourier modes of the linearized Boltzmann equation, and
identify the $k-$dependent transport coefficients from the decay
rates. This method is particularly useful if there exist
intermediate length scales in the problem \cite{gerits}, as is the
case here.

The Fourier modes of spatial fluctuations decay like
$\exp[-\zeta(k)t]$, and may be divided into {\it soft} (slowly
decaying) hydrodynamic modes, and {\it hard} (rapidly decaying)
kinetic modes. The former class has a vanishing decay rate, $\zeta
(k \to 0) \propto k^2$, in the long wave length limit, and
corresponds to a {\it conserved} density. The latter class
consists of hard kinetic modes with a decay rate $\zeta (k \to 0)
=$ constant. Generalized hydrodynamics concerns the study of soft
modes of fluctuations of locally conserved densities at {\it all}
wave numbers $k$.

It is the goal of this section to calculate the dispersion
relation of the relaxation rate $\zeta (k)$ of the soft heat mode
in our model, which corresponds to the locally conserved energy
density. The behavior of  this mode will depend on how the
wavelength of the disturbance compares with the relevant length
scales in the system.

The relevant length scales in the heat conduction model are the
macroscopic system size $L$, the inter-particle distance
$n^{-1/d}$, and the range of the interaction forces $r_c$. The
ratio $L/r_c$ controls the finite size effects, and the reduced
density $\rho=n r_c^d$ is the only dimensionless parameter that
controls the dynamics of the problem.

The basic distance to determine whether a perturbation of
wavelength $\lambda_k$ decays according to standard hydrodynamics
with constant transport coefficient is the range $r_c$, i.e. for
$\lambda_k \gg r_c$ the decay of the heat mode is $\zeta (k)
\simeq k^2 D_T$, where $D_T$ is the standard heat diffusivity. In
general however, it decays as,
\be \label{D-k}
\zeta(k) \equiv D_T (k) k^2
\ee
with a $k-$dependent heat diffusivity $D_T(k)$ that approaches
 the constant transport coefficient $D_T$ as $k \to 0$.
 As soon as the wavelength
$\lambda_k$ is comparable to $r_c$, the transport coefficient
$D_T(k)$ becomes $k$-dependent. This range of excitations is called
generalized hydrodynamics. The method to study generalized
hydrodynamics in the heat conduction model is essentially the same
as the one followed in Ref. \cite{gerits} for a lattice gas
cellular automaton model of the Van der Waals equation, or in Ref.
\cite{rip01} for standard DPD without energy conservation.

To set up the kinetic theory  we start with the linearized
Boltzmann equation. To derive it we follow the method of
Ref. \cite{mbe1}, dealing with dissipative particle dynamics for
viscous dissipation. So we start with the first equation of the
BBGKY hierarchy of the one-, two-, $\cdots n-$particle reduced
distribution functions, $f(x_1,t), f^{(2)} (x_1, x_2,t), \cdots $
with phase $x_i =\{ {\bf r}_i,\epsilon_i \}$. This is done by
integrating the $N-$particle Fokker Planck or Liouville equation,
$\partial_t \rho = L \rho$  with $L$ in \Ref{L-DB}, over the
phases $ x_2, x_3, \cdots ,x_N$. The result is,
\ba \label{k1}
&\partial_t f(x_1,t)= \int dx_2 T(12) f^{(2)}(x_1, x_2,t)&
\nn & = \alpha k_B \lambda_0 \int d{\bf r}_2 d\epsilon_2
\partial_{12} w(r_{12}) [T_1-T_2 +k_BT_1 T_2 \partial_{12}]f^{(2)}({\bf
r}_1,\epsilon_1,{\bf r}_1 -{\bf R},\epsilon_2,t)&,
\ea
where $T(12)$ is the two-particle Fokker Planck operator, defined
through  $L = \sum_{i<j} T(ij)$ in \Ref{L-DB}. Note that
$[\cdots]$ in $L$ of Eqs.\Ref{L-DB} have been replaced by
$[\cdots]$ in Eq.\Ref{k1}. This implies that the correction of
relative ${\cal O}(1/\alpha)$ to $(T_1-T_2) (1-1/\alpha)$ has been
neglected for consistency. Eq.\Ref{k1} is not a closed equation
since the time evolution of the one particle distribution function
is expressed in terms of the pair distribution function. In DPD
models with their softly repulsive interactions it is in general a
reasonable approximation to assume {\it molecular chaos}, which
expresses the statistical independence of the energy fluctuations
in different particles
\cite{mbe1}, i.e.
\be \label{k2}
f^{(2)} ({\bf r}, \epsilon, {\bf r}^{\prime}, \epsilon^{\prime},
t)   \simeq f({\bf r}, \epsilon,t)f({\bf r}^{\prime},
\epsilon^{\prime}, t).
\ee
The Fokker Planck Boltzmann (FPB) equation  for the single
particle distribution $f(x_1,t)$ is then obtained by combining
\Ref{k1} and \Ref{k2}.

We are specifically interested in studying  the decay rates of the
Fourier mode, here the heat mode, as a function of the wave number
${\bf} k$ \cite{resibois}.  So, we study the decay of small
deviations from thermal equilibrium. This can be done by
linearizing the FPB equation around the equilibrium distribution
function (\ref{fcan}), $f_0({\bf x}) = n \psi_0({\epsilon})$, i.e.
\be \label{k3}
f({\bf x}_1,t) = n \psi_0(\epsilon_1) \left[ 1 + H({\bf x}_1,t)
\right],
\ee
where $H({\bf x}_1,t)=H_1$ is a small quantity. This yields,
\be \label{k4}
\partial_t \psi_0(\epsilon_1) H_1 \simeq
n \int d{\bf x}_2 T(12) \psi_0(\epsilon_1) \psi_0(\epsilon_2)
\left(1 + {\cal P}_{12} \right) H_1,
\ee
where the permutation operator, ${\cal P}_{12}$, interchanges the
labels of the two particles {\em i.e.} ${\cal P}_{ij} h_i= h_j$.
Higher terms than first order in $H(x_1,t)$ have been neglected.
We are interested in the Fourier modes of \Ref{k4}, defined as,
\be \label{k5}
H({\bf x}, t ) = e^{-\zeta(k)t + i {\bf k} \cdot {\bf r}} h({\bf
k},\epsilon),
\ee
where ${\bf k}$ is the wave vector of the Fourier mode and
$\zeta({ k})$ its decay rate. The allowed wave numbers, $k_\alpha
= 2 \pi n_\alpha /L$ with $\alpha =x,y, \cdots $, and $n_\alpha
=0,\pm 1, \pm 2, \cdots $ are determined by the periodic boundary
conditions. Substitution of (\ref{k5}) into (\ref{k4}) yields the
eigenvalue equation,
\be \label{k6}
\left[ \zeta({ k})+\Lambda({\bf k}) \right] \psi_0 (\epsilon) h =
0,
\ee
where the operator $\Lambda ({\bf k})$ is defined as
\be \label{k7}
\Lambda({\bf k}) \psi_0(\epsilon_1) h({\bf k}, \epsilon_1) = n
\int d{\bf x}_2 T(12) \psi_0(\epsilon_1) \psi_0(\epsilon_2)
\left(1 + e^{-i{\bf k}\cdot {\bf r}_{12}}{\cal P}_{12} \right)
h({\bf k}, \epsilon_1).
\ee
As a preparation to solve (\ref{k7}) we simplify  the above
expression, by using the relation,
\be \label{k8}
T(12) \psi_0(\epsilon_1) \psi_0(\epsilon_2) {B(x_1 x_2)}  = \alpha
k^2_B\lambda_0 w(r_{12}) \partial_{12} T_1 T_2 \psi_0(\epsilon_1)
\psi_0(\epsilon_2) \partial_{12}{B(x_1 x_2)},
\ee
where $B(x_1,x_2)$ is an arbitrary function of the phases. This
simplification combined with the relations: $ \epsilon_i =\alpha
k_B T_i$ and $\omega_0= \rho \lambda_0$ gives for the collision
operator \Ref{k7},
\be \label{k9}
\Lambda({\bf k}) \psi_0(\epsilon_1) h_1= (\omega_0/\alpha) \int
d\epsilon_2 \partial_{12} \epsilon_1 \epsilon_2 \psi_0(\epsilon_1)
\psi_0(\epsilon_2)
\partial_{12} \left[1 + W(k){\cal P}_{12} \right] h_1,
\ee
where the Fourier transform of $w(R)$ is,
\be  \label{k9a}
 \int d{\bf R} \exp[-i{\bf k \cdot R}] w(R) = \widetilde{w}(k) \equiv [w]W(k).
\ee
What remains to be done is to solve the eigenvalue equation
\Ref{k6}. It is a simple matter to verify that $h(k \to
0,\epsilon) = 1+\alpha-\beta\epsilon$ is an exact eigenfunction of
$\Lambda (\bk)$, and substitution in \Ref{k9} yields the
eigenvalue,
\be \label{k10}
\zeta({ k}) = \omega_0 [1-W({k})].
\ee
For {\it small} $k$, the ${\bf k-}$expansion of $W(k)$ gives
$W(k)=1 - \frac{1}{2d}\langle R^2 \rangle_w k^2 +{\cal O}(k^4)$.
For {\it large} $k$ the term $W(k) \to 0$ due to the rapid
oscillations of $\exp[- i {\bf k \cdot R} ]$. So its limiting
behavior is,
\be \label{k11}
\zeta(k)= D_T(k) k^2 = \left\{ \begin{array}{cc} D_T k^2 - B_T k^4
+ \cdots &
\qquad \mbox{ for $kr_c \ll 1$} \\[2mm]
\omega_0   &\qquad \mbox{ for $kr_c \gg 1$}
\end{array} \right.,
\ee
where $D_T$ is the standard heat diffusivity and $B_T$ the
so-called super-Burnett coefficient. Here $D_T$ coincides with the
mean field result $D_\infty$ in  \Ref{4.11}-\Ref{k10}, and
\be \label{burnett}
B_T=  \frac{ \omega_0}{8d(d+2)} \langle R^4 \rangle_w
\ee
with $\omega_0= \rho\lambda_0$. The form $(\alpha+1-\beta \epsilon)$
of the eigen mode confirms that this mode is indeed the soft mode of
interest, corresponding to the conserved energy.

\begin{figure}[hbt]
$$\includegraphics[width=0.35\columnwidth,angle=-90]{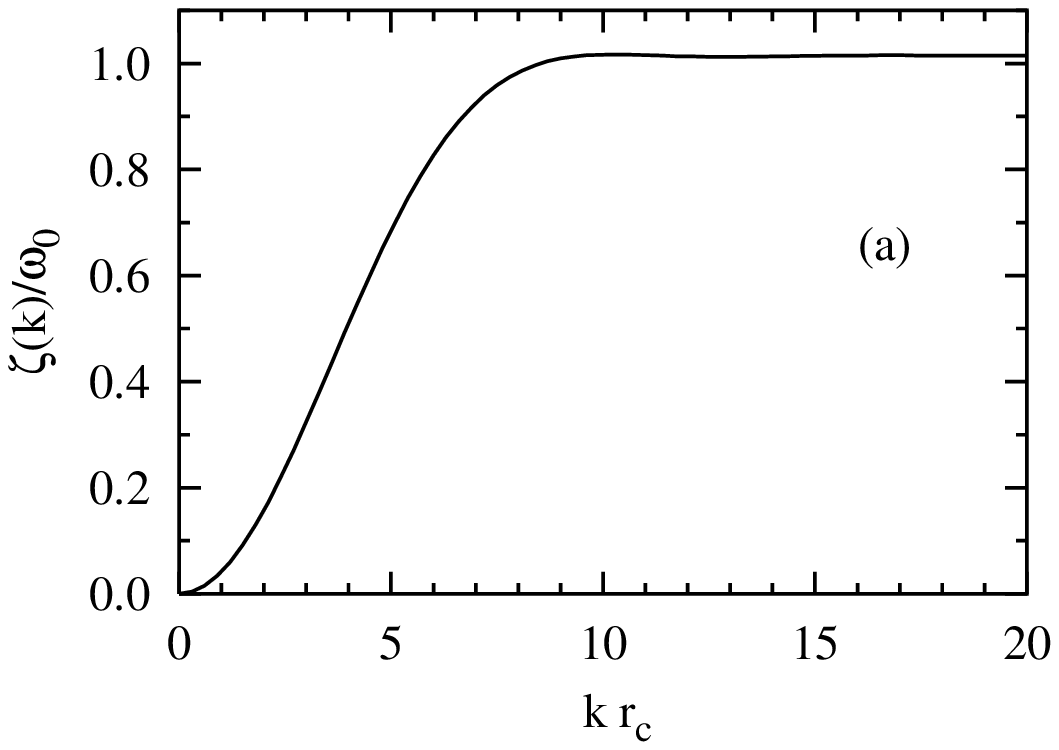}
\includegraphics[width=0.35\columnwidth,angle=-90]{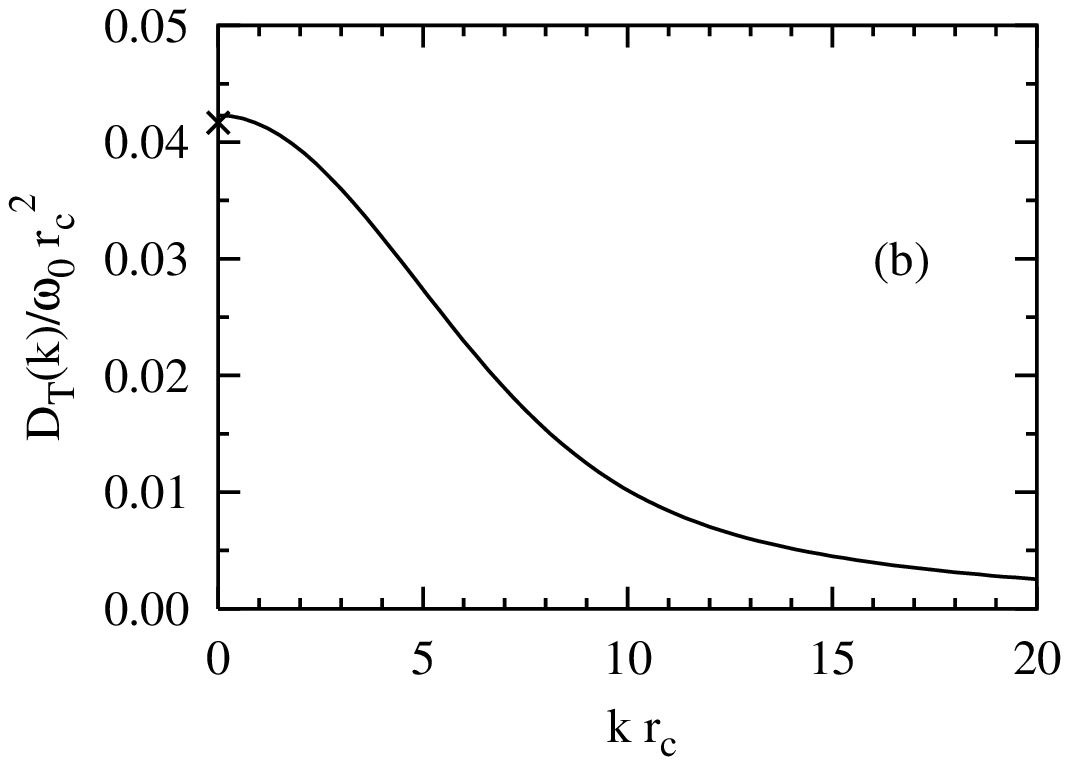}$$
\caption{ (a) Decay rate of the 3-D heat mode $\zeta (k)= D_T(k)
k^2$ in units of $\omega_0$ and (b) heat diffusivity $D_T(k) =D_T
-k^2 B_T +\cdots $ in units of $\omega_0 r_c^2$, plotted versus $k
r_c$. Solid lines in (a) and (b) are calculated with the Lucy
function in \Ref{w-lucy}, and $\alpha=100$.}
\end{figure}

At large $k$ the heat mode becomes a hard kinetic mode with a
constant decay rate, $\zeta(k \to \infty)$ =constant. This
behavior is shown in Fig.6, where the relaxation rate $\zeta (k)$
and the generalized heat diffusivity $D_T(k)$ in \Ref{k10} are
plotted versus $k r_c$ for the three-dimensional case. The Fourier
transforms $W(k)$ can be calculated by using the following
formulas,
\be \label{W-fourier}
W(k) = \left\{ \begin{array}{ll} \frac{\int_0^1  dx x w(x)
J_0(qx)}{\int_0^1  dx x w(x)}& \quad
(d=2)\\[2mm]
 \frac{\int_0^1  dx x w(x) \sin(q x)}{q\int_0^1 dx x^2 w(x)} &
\quad (d=3) \end{array} \right.
\ee
where $q=kr_c$ and $J_0(x)$ is the zeroth order Bessel function.
Note that $W(k)$ is a non-decreasing function of $k$ for the Lucy
function, and an oscillating one for the Heaviside function.

It is also worthwhile noting that the calculations of $D_T$ in
this section and those in Section IIIB give identical results,
although the structure of the calculations is very different. In
Section IIIB the heat flux is calculated in a state of local
equilibrium, which is fully factorized because conservative forces
are absent. In this section on the other hand we have derived a
Boltzmann equation for the single particle distribution function
$H(x_i,t)$, based on the molecular chaos assumption, and we solved
the eigenvalue equation for the heat mode to obtain the eigenvalue
or decay rate $\zeta (k) \simeq k^2 D_T$ at long wavelength. As the
molecular chaos assumption also neglects the spatial correlations
between colliding particles, the kinetic theory results are also mean
free results. As discussed in Section III for the formula \Ref{4.11},
also the result \Ref{k10} is only valid at high densities.

\renewcommand{\theequation}{V.\arabic{equation}}
\setcounter{section}{4} \setcounter{equation}{0}
\section{Conclusions and Prospectives}

One of the most interesting prospects of the present paper is that
the DPD solid and the Lorentz gas are relatively simple
many-particle systems that can be used to develop kinetic
equations for classical fluids that go beyond the mean field
Boltzmann equation. Moreover, they are {\it complementary } to one
another in so far as the mechanisms of transport are concerned. In
the Lorentz gas it is {\it kinetic transport}. In the DPD solid
the mechanism is {\it collisional transfer}. In classical fluids
both mechanisms are present, and the former is dominant at {\it
low} and {\it moderate} densities; the latter is dominant in dense
liquids.

In the Lorentz gas many-particle resummation techniques have been
developed that account for the collective effects of ring
collisions, i.e. sequences of dynamically correlated binary
collisions, that led to the log-$\rho$ dependence of transport
   coefficient \cite{vLW67} and to the power law long time tails in
the velocity autocorrelation function \cite{EW79}.

In the present DPD model we have firmly established  by elaborate
computer simulations, that mean field theory gives essentially
exact results for the transport coefficients at very high
densities, and that the faster-than-linear decrease of the heat
diffusivity, $ D_T(\rho) = D_\infty(\rho) \{ 1- A/\rho +
\cdots\}$, is caused by {\it local density fluctuations}. This was
done by comparing on- and off-lattice spatial configurations of
particles (see Fig.1).

Consequently, it is to be expected that applications of Effective
Medium Theory \cite{kirk73,webman81}, which is equivalent to the
self-consistent ring-kinetic equation \cite{vVE87}, and its
extensions to  classical fluids, would provide systematic methods
for calculating transport properties, starting at the {\it high
density } side of the density spectrum.

Furthermore, as the density $\rho$ decreases the heat diffusivity
rapidly decreases to zero at a threshold density $\rho_c$, below
which the heat conductivity vanishes. The existence of a heat
conduction threshold $\rho_c$ is explained as a dynamic
percolation phenomenon, and identified with bond percolation on a
random proximity network. The dynamics on this discrete network is
different from diffusion on the continuum percolation structures,
although the geometrical connectivity properties of the discrete
and continuous percolating cluster are the same. The threshold is
identified with the two-dimensional percolation threshold $\rho_c
= 1.43629$ \cite{ziff-3d} of continuum percolation of overlapping
spheres, and its value agrees well with the known conductivity
threshold $\rho_c$ in Fig.3a. So the simulation values for the heat
conductivity agree both at high and low density with our
theoretical analysis of the heat conductivity. Our older
three-dimensional simulation data for the heat conductivity of
Ref. \cite{Ripoll-thesis}, shown in Fig.1, are not inconsistent
with the existence of a conductivity threshold $\rho_c$ at the 3-D
percolation threshold $0.65296$ \cite{ziff-3d}, but simulation
data are lacking close to the percolation point, and presumably
show strong finite size effects, caused by the small systems used
in the simulations.

We have further extended the kinetic theory to the regime of
generalized hydrodynamics by studying the wave number dependent
decay rates of the Fourier modes of the temperature fluctuations.
The decay rate $\zeta (k) = k^2 D_T(k)$ of these modes depends
strongly on how the probing wavelength (size of colloidal
particles, polymers, pores....) compares to the range of the DPD
forces. So, there are several possibilities for applying the
generalized hydrodynamics results, apart from the applications,
already discussed in Section IV, in mode coupling theories, and in
analyzing finite size effects where the discreteness of the
allowed ${\bf k}-$values has to be taken into account. In the
limit of long wave lengths the heat diffusivity $D_T(k \to 0)$
reaches the constant value given by the standard Chapman Enskog
theory. When the wave length of the perturbation is of the same
order of magnitude as the range $r_c$ of the forces, the heat
diffusivity, predicted by the generalized hydrodynamics, decreases
significantly below its long wave length value. Here we also
mention the application of our $k-$dependent transport
coefficients in {\it smoothed} DPD \cite{SDPD}. The goal of such
methods is to discretize macroscopic nonlinear partial
differential equations -- here Fourier's law for heat diffusivity
-- and to solve them with molecular dynamics codes (see
Ref.\cite{esp01revenga}).  The finite size effects, discussed in
Section IIIB, are in smoothed DPD, as well as in the related
Smooth Particle Hydrodynamics \cite{Ripoll-thesis}, measures for
controlling the discretization errors. The authors of Ref.
\cite{SDPD} have measured and analyzed the decay rates $\zeta (k)$
of sinusoidal temperature profiles in our heat conduction model,
from which  the values of $D_T(\bk)$ are extracted, and compared
with our results for $D_T(k)$, as presented in Section IV.

\section*{Acknowledgments}
The authors thank Pep Espa\~nol for useful discussions. M.H.E.
thanks R.M. Ziff for his extensive explanations about continuum
percolation problems, and J. Machta for helpful correspondence.
M.R. thanks Gerrit Vliegenthart for valuable discussions. M.R.
also acknowledges financial support from the Ministerio de Ciencia
y Tecnolog{\'{\i}}a under the project BFM2001-0290 and the German
Research Foundation (DFG) within the SFB TR6.

\appendix

\renewcommand{\theequation}{A.\arabic{equation}}
\setcounter{equation}{0}
\section{H-theorem and equilibrium State}

 In this appendix we  prove an ${\cal H}$ theorem, and analyze
the equilibrium distribution. We show that the function ${\cal H}$
is a Lyapunov functional with $\partial_t{\cal H} \leq 0$, and
investigate the implications of this result for the equilibrium
solution of the Fokker Planck equation.

We consider the following functional of the $N$-particle
distribution function $\rho({\bf X})$,
\be \label{hfun}
{\cal H} [\rho] = \int d {\bf X} \left[ \ln \rho({\bf X}) -
\frac{S({\bf X})}{k_B} \right] \rho({\bf X}),
\ee where $S({\bf X})=\sum_i^N s(\epsilon_i)$ and $s(\epsilon_i)$ is the one
particle entropy function, and $-{\cal H}$ is the total entropy of
the $N-$particle system. Similar results have been obtained in
Refs.\cite{mbe1,marsh98,Ripoll-thesis}.

The time derivative of the functional in (\ref{hfun}) is given
through the Fokker Planck equation,
\ba \label{FPE.lyap}
&\partial_t {\cal H} = \int d {\bf X} \left[ \ln \rho({\bf X}) -
\frac{S({\bf X})}{k_B}\right] \partial_t \rho({\bf X}) = \int d
{\bf X} \rho({\bf X}) L^{\dagger} \left[ \ln \rho({\bf X}) -
\frac{S({\bf X})}{k_B} \right]&
\nn & =-\int d\bX \rho (\bX) \sum_{i<j}A_{ij}B_{ij},&
\ea
where we have performed partial integrations, and introduced the
symbols,
\ba \label{Aij}
A_{ij} & =& \partial_{ij} \left[\ln \rho({\bf X}) - \frac{S({\bf
X})}{k_B} \right]= \partial_{ij} \ln \rho - \frac{1}{k_B T_i} +
\frac{1}{k_B T_j}
\nn B_{ij} &=& \lambda(ij)(T_i-T_j) +\half \partial_{ij} [a^2(ij)
\rho]
\nn&=& \half a^2(ij) \rho \left\{\frac{2\lambda(ij)
T_iT_j}{a^2(ij)}\left( \frac{1}{T_j} -\frac{1}{T_i}  \right) +
\partial_{ij} \ln [a^2(ij)\rho] \right\}.
\ea
The strategy is to make the factor $\{ \cdots\}$ in $B_{ij}$ equal
to $A_{ij}$, yielding,
\be \label{ht.hc}
\partial_t {\cal H} =
- \int d {\bf X} \rho({\bf X}) \sum_{i < j} \half a^2(ij) (
A_{ij})^2 \leq 0,
\ee
which guarantees that $\partial_t {\cal H}$ in \Ref{FPE.lyap} is
{\it non-decreasing}. This is the desired ${\cal H}$-theorem.
There are two possibilities to realize this. The {\it first} one
is to choose,
\be \label{db.k}
a^2(ij) = 2k_B \lambda(ij) T_iT_j \quad \mbox{and} \quad
\partial_{ij} a^2(ij) =0 \quad (\forall \{i,j\}).
\ee
These conditions are referred to as {\it Detailed Balance}
conditions. A solution of the last equation is,
\be \label{DB-1}
a^2(ij) = 2k_B \lambda(ij) T_iT_j \equiv \kappa (ij) =2 k_B
w(r_{ij})F(T_i+T_j).
\ee
Here $w(r)$ is the range function defined in \Ref{w-range}, which
implies $w_s(r) =\sqrt{w(r)}$, and $F(x)$ is some positive
function of $x$, e.g. $F(x) = \kappa_0 x^{2n} \; (n=0,1,2, ...)$.
This is the solution, used in
\cite{marsh98,ripoll98,Ripoll-thesis}. With the help of
\Ref{w-range} and \Ref{DB-1} the temperature relaxation equation
would take the form,
\be \label{non-thermo}
 \frac{dT_1}{dt} = \frac{\kappa_0}{\alpha k_B T_1T_2}(T_1+T_2)^{2n}
 (T_2-T_1)
 \ee
 with $n=0,1,2 ..$. Although mathematically acceptable this
 temperature relaxation equation is not in agreement with {\it
 irreversible thermodynamics}.

 Next we discuss the {\it second} set of solutions. To do so we
 drop the second requirement in \Ref{db.k}, and write the term
 $\{ ...\}$ in \Ref{Aij} as,
 \ba \label{neglect}
 \{...\} &=& \frac{1}{k_B} \left( \frac{1}{T_j} - \frac{1}{T_i} \right)
+\partial_{ij} \ln [ T_i T_j \lambda(ij) \rho ]
\nn &=&\frac{1}{k_B} \left(1 -\frac{1}{\alpha}\right)  \left( \frac{1}{T_j} -
\frac{1}{T_i}\right)+\partial_{ij} \ln \lambda (ij) +
\partial_{ij}\ln \rho.
\ea
As explained below \Ref{ecstate} $\alpha$ is a {\it large} number,
measuring the number of internal degrees of freedom of a DPD
particle, and $(1-1/\alpha)$ should be replaced by 1 for
consistency to leading order in $1/\alpha$. Now the expression $\{
...\}$ in \Ref{neglect} can be made equal to $A_{ij}$ in
\Ref{FPE.lyap} by choosing $\lambda(ij)$ {\it independent} of the
internal energies of the interacting particles, and in {\it
agreement} with  the laws of irreversible thermodynamics
\Ref{irr-thermo} and \Ref{lambda-const}. Then the second set of
{\it Detailed Balance} conditions becomes,
\be \label{DB-2}
a^2(ij) = 2k_B \lambda(ij) T_i T_j \quad \mbox{and} \quad \lambda
(ij) =\alpha k_B\lambda_0 w(r_{ij}).
\ee
We also quote for later reference that both sets of Detailed
Balance conditions \Ref{db.k} and \Ref{DB-2} guarantee the
commutation relation,
\be \label{commute}
a^2(ij) \partial_{ij} = \partial_{ij} a^2(ij) +{\cal O}(1/\alpha).
\ee

The ${\cal H}$-function in \Ref{ht.hc} keeps decreasing and
reaches a minimum, if and only if the partial differential
equations, $A_{ij}=0$, are satisfied for all $\{i,j\}$. The
solution of these differential equations determines the
equilibrium distribution. As the particles are point particles,
the $N-$particle distribution factorizes,
\be \label{rofact}
\rho_{eq} ({\bf X}) = \frac{1}{Z} \prod_i \psi_0 (\epsilon_i).
\ee Combination of these differential equations,
with \Ref{rofact} yields for all possible
pairs $(ij)$,
\be \label{eq10a}
\frac{\partial \ln \psi_0}{\partial \epsilon_i} - \frac{1}{k_B
T_i} = \frac{\partial \ln \psi_0}{\partial \epsilon_j} -
\frac{1}{k_B T_j} = -
\beta,
\ee where $\beta$ is a constant with dimensions of an inverse energy.
By using the relation, $\partial s(\epsilon_i)/\partial\epsilon_i
= 1 / T_i$, integration of (\ref{eq10a}) yields
\be \label{eq.cond1}
\psi_0(\epsilon_i) =\frac{1}{z(\beta)} g(\epsilon_i)\exp[-\beta
\epsilon_i] =\frac{1}{z(\beta)}\exp[k_B^{-1} s(\epsilon_i) -
\beta \epsilon_i],
\ee where the factor $g(\epsilon_i) \equiv \exp[k_B^{-1}
s(\epsilon_i)] \propto \epsilon_i^\alpha$ is the {\em degeneracy
factor} or the number of internal states having energy
$\epsilon_i$.  In a mesoscopic picture  $g(\epsilon_i)$  is the
number of internal states of the mesoscopic particle $i$ having
energy $\epsilon_i$. The normalization factor $z(\beta)$ is,
\begin{equation} \label{part1}
z(\beta) = \int_0^\infty d\epsilon \exp
[k_B^{-1}s(\epsilon)-\beta\epsilon ].
\end{equation}
This factor corresponds to the partition function of a single
mesoscopic particle.

For the DPD solid, where the entropy is given in (\ref{def.s}),
the one particle equilibrium distribution function becomes,
\begin{equation} \label{a11}
\psi_0(\epsilon) =
\frac{\beta}{\Gamma(\alpha+1)}\left(\beta\epsilon\right)^\alpha
\exp[-\beta\epsilon],
\end{equation}
where $\Gamma(x)$ is the gamma function. For later use we also quote the moments of $\psi_0$,
\be  \label{pe.moments}
\langle \epsilon^n \rangle = \int d\epsilon \psi_0(\epsilon)
\epsilon^n = \frac{\Gamma(\alpha+n+1)}{\Gamma(\alpha+1)}
\beta^{-n},
\ee
in particular
\be \label{e-fluct}
\av{(\delta\epsilon_i)^2}=
\av{\epsilon_i^2}-\av{\epsilon_i}^2=(\alpha+1)/\beta^2 \simeq
\alpha/\beta^2.
\ee
The parameter $\beta$ is related to the total energy of the system
$E$ through the relation,
\be \label{beta.ener}
\frac{E}{N} =-\frac{\partial \ln z(\beta)}{\partial \beta} =
\frac{(\alpha+1)}{\beta} \simeq \alpha/\beta.
\ee
The last equality has been calculated by considering the entropy
function (\ref{def.s}). Furthermore, the following average can be
calculated
\be \label{temp1}
\left\langle \frac{1}{T_i}\right\rangle_\beta  =
\frac{1}{z(\beta)} \int_0^\infty d\epsilon \frac{\partial
s(\epsilon)}{\partial \epsilon}
\exp\left[k_B^{-1}s(\epsilon)-\beta\epsilon \right] = k_B\beta.
\ee
Note that this relation is valid for a general entropy
function. It allows one to define the macroscopic temperature $T$
as $T^{-1}=\langle T^{-1}_i\rangle$ where $\beta=1/k_B T$ is the
inverse macroscopic temperature in thermal equilibrium. For the
special choice of (\ref{def.s}), it is also interesting to point
out that, $\langle T_i \rangle = [(\alpha+1)/\alpha]\; T \simeq T$,
this is the relation between the macroscopic temperature and the
average temperature.  As discussed below (\ref{ecstate}), $\alpha$
scales like the number of internal degrees of freedom and therefore
$\alpha \gg 1$.

The discussion above deals with fluctuations in the {\it single
particle} energies, calculated in the canonical ensemble. In
\cite{Ripoll-thesis,ripoll98} it was shown that such fluctuations,
calculated in the micro-canonical ensemble, give the same results,
provided $\alpha$ is large. This condition is always satisfied as DPD
particles are mesoscopic objects.


\end{document}